\documentclass[aps,prb,twocolumn,showpacs]{revtex4}
\usepackage{epsfig}
\usepackage{times}
\usepackage{amsmath}
\begin{document}

\title{Spatial coherence resonance on diffusive and small-world networks of Hodgkin-Huxley neurons}

\author{Xiaojuan Sun}
\email{sxiaojuan@ss.buaa.edu.cn}
\affiliation{School of Science, Beihang University, Beijing 100083, China and Institute of Physics, University of Potsdam, PF 601553, 14415 Potsdam, Germany}

\author{Matja{\v z} Perc}
\email[Electronic address (corr.): ]{matjaz.perc@uni-mb.si}
\affiliation{Department of Physics, Faculty of Natural Sciences and Mathematics, University of \\ Maribor, Koro{\v s}ka cesta 160, SI-2000 Maribor, Slovenia}

\author{Qishao Lu}
\email{qishaolu@hotmail.com}
\affiliation{School of Science, Beihang University, Beijing 100083, China}

\author{J\"urgen Kurths}
\email{jkurths@agnld.uni-potsdam.de}
\affiliation{Institute of Physics, University of Potsdam, PF 601553, 14415 Potsdam, Germany}

\begin{abstract}
Spatial coherence resonance in a spatially extended system that is locally modeled by Hodgkin-Huxley (HH) neurons is studied in this paper. We focus on the ability of additive temporally and spatially uncorrelated Gaussian noise to extract a particular spatial frequency of excitatory waves in the medium, whereby examining also the impact of diffusive and small-world network topology determining the interactions amongst coupled HH neurons. We show that there exists an intermediate noise intensity that is able to extract a characteristic spatial frequency of the system in a resonant manner provided the latter is diffusively coupled, thus indicating the existence of spatial coherence resonance. However, as the diffusive topology of the medium is relaxed via the introduction of shortcut links introducing small-world properties amongst coupled HH neurons, the ability of additive Gaussian noise to evoke ordered excitatory waves deteriorates rather spectacularly, leading to the decoherence of the spatial dynamics and with it related absence of spatial coherence resonance. In particular, already a minute fraction of shortcut links suffices to substantially disrupt coherent pattern formation in the examined system.
\end{abstract}


\maketitle

\textbf{Nontrivial effects of noise on nonlinear dynamics have been a vibrant topic for many years. It is thoroughly documented and established that noise can play a constructive role in different types of nonlinear dynamical systems. Stochastic and coherence resonance are just two, perhaps most prominent, examples of this fact. The notion of coherence resonance is particularly inspiring as it messages that an appropriate intensity of noise alone is sufficient to evoke ordered temporal responses of a nonlinear dynamical system. Nowadays, however, effects of noise on spatially extended systems have gradually slipped into the focus of many scientists working in diverse fields of research, consequently spawning the need to investigate whether phenomena observed previously for isolated dynamical system can also be observed, at least conceptually similar, if the latter are coupled. Indeed, it has been shown that the coherence resonance phenomenon originally reported for dynamical systems evolving only in time can also be observed in spatially extended systems that are locally described by excitable nonlinear dynamics. Importantly thereby is the fact that the previously studied order in the temporal dynamics has been "replaced" by the order of the noise-induced spatial dynamics, which mostly manifests as propagating waves of excitatory events throughout the spatial grid. Presently, we aim to extend the scope of spatial coherence resonance by confirming its possibility also in models of neuronal dynamics, in particular by employing as the constitutive unit of the spatially extended system the renowned HH model. We show that while the diffusive connectivity of coupled neurons warrants the observation of noise-induced pattern formation and with it related spatial coherence resonance, the small-world topology is not an appropriate medium for such observations. More precisely, even a minute fraction of shortcut links amongst distant neurons prohibits noise-induced waves to be ordered, and hence also precludes the observation of spatial coherence resonance. Since the present study is set-up around a comprehensive HH model of neuronal dynamics, the presented results should prove valuable not just from the purely theoretical point of view, but hopefully also from the experimental point of view, especially by shedding light into the functioning of neural tissue.}

\section{Introduction}
Randomness is a common feature in the real world. It is well known that noise can play a constructive role in different nonlinear dynamical systems, such as optical devices, \cite{Fioretti_1993} electronic circuits, \cite{Zhou_1990} or neural tissue. \cite{Ullner_2003} Stochastic and coherence resonance are two prominent examples of such constructive effects that rely on the influence of noise on nonlinear systems. \cite{Gam_1998, Lind_2004} Remarkably, by the coherence resonance \cite{Hu_1993,Rap_1994,Pik_1997} an enhanced ordered behavior results solely from the introduction of noise in the absence of additional weak deterministic signals that are otherwise a standard ingredient by the observation of stochastic resonance. \cite{Gam_1998} The phenomenon of coherence resonance has been studied extensively also in non-identical coupled neurons \cite{Zhou_2001, Kurths_2003} as well as one-dimensional arrays of nonlinear dynamical systems in general. \cite{Wio_1995, Han_1999, Neiman_1999} Zhou and Kurths \cite{Zhou_2002} showed, for example, that in a weakly coupled region noise could induce spatio-temporal coherence resonance but reduced the degree of synchronization by stronger couplings.

Recently, many authors have shifted their interests to the influences of noise on two-dimensional spatially extended systems, and to their spatial rather than temporal dynamics in particular. \cite{Gar_2007} Spatio-temporal stochastic resonance has been reported first in, \cite{Jung_1995} while the spatial coherence resonance has also been studied in many different types of spatially extended systems, e.g. in chlorine dioxide-iodine-malonic reactions, \cite{Car_2004} Rulkov maps, \cite{Perc_2007} or excitable biochemical media. \cite{Gosak_2007} More precisely, Carrillo et al. \cite{Car_2004} demonstrated that spatial coherence resonance could be evoked close to pattern-forming instabilities, mimicking one of the mechanisms of standard temporal coherence resonance, while Perc et al. \cite{Perc_2007, Gosak_2007} showed that in locally discrete excitable spatially extended systems and excitable biochemical media additive or internal noise could also lead to the observation of spatial coherence resonance. Importantly, however, there is still a lack of comprehensive studies investigating whether the above results concerning spatial coherence resonance can be obtained also in more complex spatially extended systems, particularly such that are locally modeled by realistic nonlinear dynamical systems that faithfully describe a real-life biological process, as is for example the neuronal activity.

In this paper, we would like to address the above-described void and extend the scope of spatial coherence resonance by examining the possibilities of its existence in a spatially extended system that is locally modeled by excitable HH neurons. \cite{Hod_1952} More precisely, we analyze spatial frequency spectra of the examined medium in dependence on different levels of additive noise and topologies of the interaction network constituting the couplings amongst the HH neurons. By calculating the average spatial structure function, we present conclusive evidences for spatial coherence resonance provided the neurons are diffusively coupled. In particular, we show that then there exists an optimal level of additive noise for which a particular spatial frequency of excitatory events in the medium is best pronounced. We emphasize that thereby no additional deterministic inputs are introduced to the system, and the latter is locally initiated from steady state initial conditions. In contrast with results obtained when employing diffusive nearest-neighbor interactions, however, we find that small-world interaction networks \cite{Watts_1998, Watts_1999} fail to sustain coherent patterns of spatial noise-induced excitations, and thus do not warrant the observation of spatial coherence resonance. More precisely, even a minute fraction of rewired links, connecting two nearest neighbors in case of diffusive coupling, heavily disrupts coherent pattern formation in the medium and induces decoherence of excitatory waves. We thus reveal the impact of additive noise and small-world topology on an ensemble of HH neurons, hence providing interesting theoretical insights into to the functioning of neural tissue.

The paper is structured as follows. Section II is devoted to the description of the employed spatially extended system and the HH mathematical model as its main ingredient. Sections III and IV feature evidences for spatial coherence resonance on diffusive grids and decoherence of pattern formation on small-world networks, respectively, while in the last Section we summarize the results and outline biological implications of our findings.

\section{Mathematical model}
In 1952, Hodgkin and Huxley presented a mathematical model to predict the quantitative behavior of an isolated squid giant axon. \cite{Hod_1952} Since then, the model has become a paradigm for mathematically describing neuron functioning, and many authors have studied its nonlinear dynamics. In particular, it has been discovered that with the change of an external current injected into the cell membrane, different bifurcations can occur rendering a stable HH neuron oscillatory, bursting, or even chaotic. \cite{Lee_1998, Rinz_1978, Troy_1978, Has_2000, Hide_2000, Wang_2005} Moreover, the dynamics of coupled HH neurons has also been studied quite extensively in the past. \cite{Hansel_1993, Mao_2005, Wang_2007} In this paper, we employ the HH model as the constitutive unit of a spatially extended system, and study the effects of different intensities of additive noise and topologies of the interaction network on the spatial dynamics of excitatory waves. The equations of a single HH model are given as follows:
\begin{subequations}
\label{eq:eq1}
\begin{eqnarray}
CdV/dt=&&-g_{Na}m^3h(V-V_{Na})-g_Kn^4(V-V_K) \nonumber \\*
&&-g_L(V-V_L)+I_{ext},
\end{eqnarray}
\begin{equation}
dm/dt=\alpha_{m}(1-m)-\beta_{m}m,
\end{equation}
\begin{equation}
dh/dt=\alpha_{h}(1-h)-\beta_{h}h,
\end{equation}
\begin{equation}
dn/dt=\alpha_{n}(1-n)-\beta_{n}n.
\end{equation}
\end{subequations}
In Eq.~(\ref{eq:eq1}) $V$ is the membrane potential of the neuron, $I_{ext}$ is the external stimulus current, whereas $m$, $h$, and $n$ are the gating variables describing the ionic transport through the membrane. Moreover, the constants $g_{Na}$, $g_K$, and $g_L$ are maximal conductances for ion and leakage channels, while $V_{Na}$, $V_K$, and $V_L$ are the corresponding reversal potentials. The functions $\alpha_{m}$, $\alpha_{h}$, $\alpha_{n}$, $\beta_{m}$, $\beta_{h}$, and $\beta_{n}$ are defined as:
\begin{subequations}
\label{eq:eq2}
\begin{equation}
\alpha_{m}=0.1(V+40)/ \{ 1- \exp[-(V+40)/10] \},
\end{equation}
\begin{equation}
\alpha_{h}=0.07 \exp[-(V+65)/20],
\end{equation}
\begin{equation}
\alpha_{n}=0.01(V+55)/ \{ 1- \exp[-(V+55)/10] \},
\end{equation}
\begin{equation}
\beta_{m}=4.0 \exp[-(V+65)/18],
\end{equation}
\begin{equation}
\beta_{h}= \{1 + \exp[-(V+35)/10] \}^{-1},
\end{equation}
\begin{equation}
\beta_{n}=0.125 \exp[-(V+65)/80].
\end{equation}
\end{subequations}
In the literature one may find different forms of the HH equations, mainly depending on the zeros of the potential. Here we choose the form so that the membrane potential at rest is $V \approx -61{\rm mV}$. Then according to the studies of Hodgkin and Huxley, \cite{Hod_1952} the parameter values in Eq.~(\ref{eq:eq1}) are as follows: $C=1{\rm \mu F/cm^2}, g_{Na}=120{\rm ms/cm^2}, g_K=36{\rm ms/cm^2}, g_L=0.3{\rm ms/cm^2}, V_{Na}=50{\rm mV}, V_K=-77{\rm mV}$, and $V_L=-54.4{\rm mV}$.

First, we outline some properties of a single HH neuron evoked by
the above parameters dependent on the external stimulus $I_{ext}$. \cite{Lee_1998} When $I_{ext}<I_0 \approx 6.2$ there exists only a globally stable fixed point. The birth of stable and unstable limit cycles occurs at $I_{ext}=I_0 \approx 6.2$ via a saddle-node bifurcation. For $I_0<I_{ext}<I_1 \approx 9.8$ there coexist a stable fixed point, a stable limit cycle, and an unstable limit cycle. The unstable limit cycle constitutes the boundary separating the attractive basins corresponding to the fixed point and the limit cycle, respectively. For a more precise description of the bifurcation structure of the HH model, in dependence also on other system parameters, we refer to. \cite{Lee_1998, Rinz_1978, Troy_1978, Has_2000, Hide_2000, Wang_2005}

\begin{figure}
\centerline{\epsfig{file=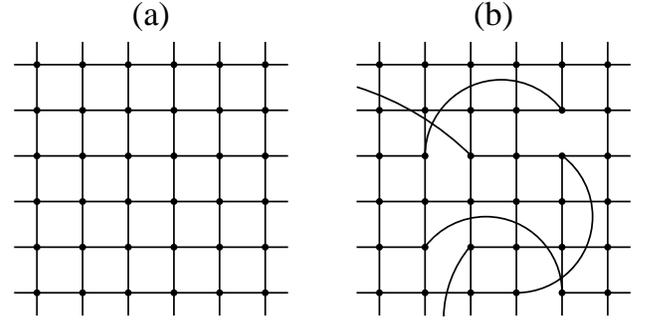,width=8.5cm}}
\caption{\label{fig:fig0}Examples of considered network topologies. For clarity only a $6 \times 6$ excerpt of the whole network is presented in each panel. (a) Diffusively coupled network characterized by $q=0$. Each vertex is directly connected only to its four nearest neighbors, hence having connectivity $z=4$. (b) Realization of small-world topology via random rewiring of a certain fraction $q$ of links, constrained only by the requirement that the initial connectivity $z=4$ of each unit must be preserved.}
\end{figure}

Based on the original HH model given by Eqs.~(\ref{eq:eq1}) and (\ref{eq:eq2}), we introduce the spatially extend system
\begin{subequations}
\label{eq:eq3}
\begin{eqnarray}
CdV_{i,j}/dt=&&-g_{Na}m^3_{i,j}h_{i,j}(V_{i,j}-V_{Na}) \nonumber \\*
&&-g_Kn^4_{i,j}(V_{i,j}-V_K)+\sigma\xi_{i,j}(t) \nonumber \\*
&&-g_L(V_{i,j}-V_L)+I_{ext} \nonumber \\*
&&+D \mathop {\sum}\limits_{k,l}\varepsilon_{i,j,k,l}(V_{k,l}-V_{i,j}),
\end{eqnarray}
\begin{equation}
dm_{i,j}/dt=\alpha_{m_{i,j}}(1-m_{i,j})-\beta_{m_{i,j}}m_{i,j},
\end{equation}
\begin{equation}
dh_{i,j}/dt=\alpha_{h_{i,j}}(1-h_{i,j})-\beta_{h_{i,j}}h_{i,j},
\end{equation}
\begin{equation}
dn_{i,j}/dt=\alpha_{n_{i,j}}(1-n_{i,j})-\beta_{n_{i,j}}n_{i,j},
\end{equation}
\end{subequations}
where subscripts $i,j=1, \ldots , N$ denote each of the $N \times N$ coupled HH neurons.
The sum in Eq.~(\ref{eq:eq3}) runs over all lattice sites whereby $\varepsilon_{i,j,k,l}=1$ if the site $(k,l)$ is coupled to $(i,j)$ and $\varepsilon_{i,j,k,l}=0$ otherwise. If the fraction of randomly rewired links $q$, constituting the small-world topology if $0 <  q \ll 1$, equals zero,
$\varepsilon_{i,j,k,l}=1$ only if $(k,l)$ is one of the four nearest neighbors of site $(i,j)$ on a regular two-dimensional mesh. Thereby, we obtain a diffusively coupled network of HH neurons as depicted in Fig.~\ref{fig:fig0}(a), which will form the backbone for the study of noise-induced spatial dynamics in Section III. If $q>0$, however, the corresponding fraction of nearest-neighbor links is randomly rewired via the analogy with the creation of Watts-Strogatz small-world networks, \cite{Watts_1998, Watts_1999} whereby we preserve the initial connectivity $z=4$ (as by the diffusive nearest-neighbor coupling) of each HH neuron as depicted in Fig.~\ref{fig:fig0}(b) to focus explicitly on the effect of network randomness rather than possible effects originating from different numbers of inputs per neuron. The precise procedure for generating such small-world networks is given in, \cite{szabo_jpa04} while results obtained when using them as interaction networks for the noise driven HH spatially extended system will be presented in Section IV. Importantly, each interaction network was generated at the beginning of a particular simulation and was held fixed, and if necessary, results presented below were averaged over $30$ different realizations of the interaction network by each $q$. Turning back to Eq.~(\ref{eq:eq3}), $\sigma$ is the main control parameter to be varied in this study, denoting the standard deviation of additive uncorrelated Gaussian noise $\xi_{i,j}$ that satisfies $<\xi_{i,j}(t)>=0$ and $<\xi_{i,j}(t)\xi_{m,n}(t^{'})>=\delta(t-t^{'})\delta_{i,m}\delta_{j,n}$. Moreover, $D=0.35$ is the presently employed coupling strength, while all other parameters are taken the same as mentioned by the description of a single HH neuron. In the present study the periodic boundary condition is used, namely: $V_{0,j}=V_{N,j}$, $V_{N+1,j}=V_{1,j}$, $V_{i,0}=V_{i,N}$, and $V_{i,N+1}=V_{i,1}$, and the system size is $N=128$ in each spatial dimension of the two-dimensional grid. In order to study explicitly the impact of noise on the spatial dynamics of the system, we take the external current $I_{ext}=6.1$ (in accordance with the above outlined properties of a single HH neuron assuring a unique stable steady state solution of each coupled unit) and initiate all units from the excitable steady state $(-61.198, 0.08199, 0.46014, 0.37727)$. Thus, without the addition of noise the medium would remain quiescent forever, and consequently the dynamics of below-discussed spatial patterns is evoked solely by additive Gaussian noise.

\section{Spatial coherence resonance by diffusive coupling}
\begin{figure*}
\centering
\includegraphics[width=14cm]{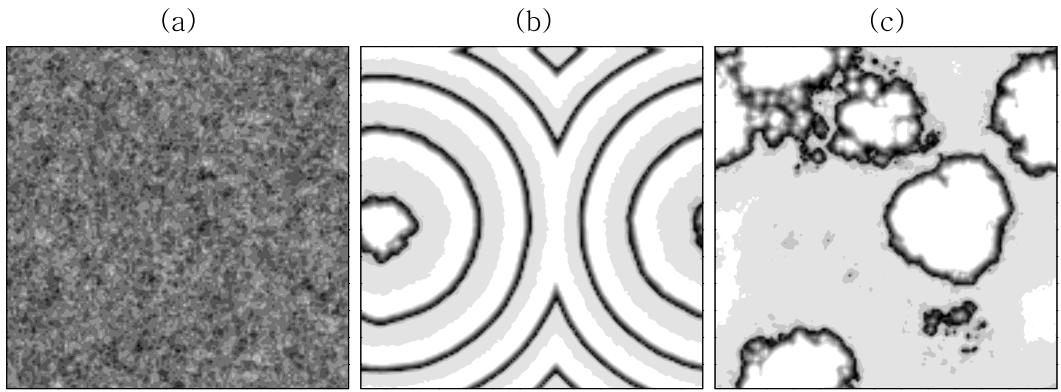}
\caption{Characteristic snapshots of the spatial profile of $V_{i,j}$ obtained for (a) $\sigma=1.1$, (b) $\sigma=1.3$ , and (c) $\sigma=1.9$. For the intermediate value of $\sigma$ the spatial dynamics is clearly most coherent, exhibiting ordered circular waves propagating through the spatial grid.}
\label{fig:fig1}
\end{figure*}

As announced, in this section we take $q=0$ whereby the coupling term $D \mathop {\sum}\varepsilon_{i,j,k,l}(V_{k,l}-V_{i,j})$ in Eq.~(\ref{eq:eq3}) reads as $D(V_{i-1,j}+V_{i+1,j}+V_{i,j-1}+V_{i,j+1}-4V_{i,j})$. We start by visually examining three snapshots of the spatial profile of $V_{i,j}$ obtained by three different values of $\sigma$, as presented in Fig.~\ref{fig:fig1}. Evidently, there exists an intermediate value of $\sigma$, at which nicely ordered circular excitatory waves propagate through the spatial domain [Fig.~\ref{fig:fig1}(b)]. On the other hand, smaller or larger values of $\sigma$ clearly fail to have the same effect evoking either only small-amplitude deviations from the steady state [Fig.~\ref{fig:fig1}(a)] or rather violent and uncorrelated excitations throughout the spatial grid [Fig.~\ref{fig:fig1}(c)], respectively. Hence, the visual inspection of the snapshots already gives some indication for a typical coherence resonance scenario for the spatial dynamics of the studied HH medium.

\begin{figure*}
\centering
\includegraphics[width=15.5cm]{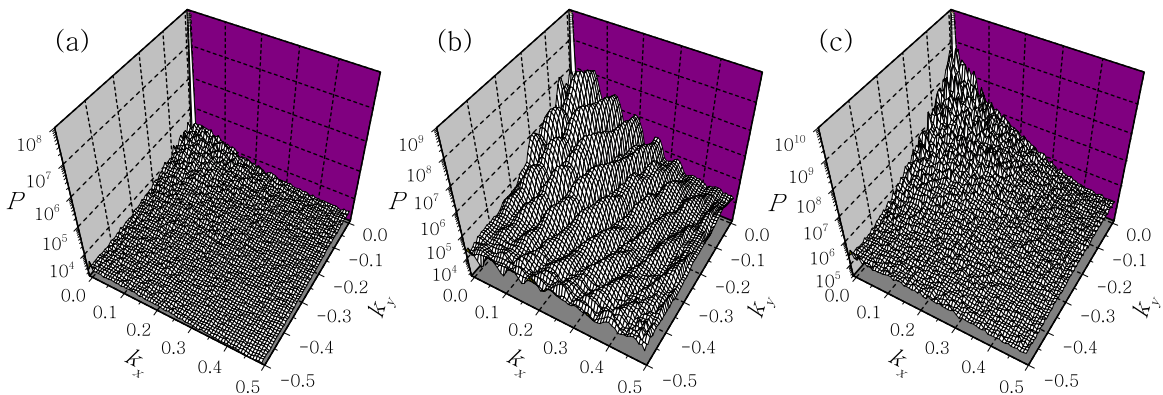}
\caption{The spatial structure function of $V_{i,j}$ obtained for (a) $\sigma=1.1$, (b) $\sigma=1.3$, and (c) $\sigma=1.9$. For the intermediate value of $\sigma$ the characteristic waterfall-like outlay of $P(k_x,k_y)$ is evident, indicating the existence of a preferred spatial frequency of noise-induced excitatory events in the medium. Note that only an excerpt of the whole $P(k_x,k_y)$ plane is shown in all panels.}
\label{fig:fig2}
\end{figure*}

To make the above observations quantitative, we calculate the spatial structure function of the $V_{i,j}$ field according to
\begin{equation}
P(k_x,k_y)=\langle H^2(k_x,k_y)\rangle,
\label{eq:eq4}
\end{equation}
where $H(k_x,k_y)$ is the spatial Fourier transform of the $V_{i,j}$ field at a particular time $t$, and $\langle ... \rangle$ is the ensemble average over noise realizations. The results from numerical simulations for different $\sigma$ are presented in Fig.~\ref{fig:fig2}. The three depicted panels correspond to the same values of $\sigma$ as used already in Fig.~\ref{fig:fig1}. Indeed, the results in Fig.~\ref{fig:fig2} fully support our visual assessments, as it can be observed nicely that for small and large noise levels [Fig.~\ref{fig:fig2}(a) and (c), respectively] the presented spectra show no particularly expressed spatial frequency, although a close examination of $P(k_x,k_y)$ at $\sigma=1.9$ still reveals some remnants of structure formation in the system. Only for intermediate levels of noise the spatial structure function develops several well-expressed circularly symmetric rings, indicating the existence of a preferred spatial frequency induced by additive Gaussian noise. As the noise level is increased, random fluctuations start to dominate the spatial dynamics and thus, similar as by small noise levels, the characteristic waterfall-like outlay of $P(k_x,k_y)$ vanishes and no preferred spatial frequency can be inferred.

\begin{figure*}
\centering
\includegraphics[width=14cm]{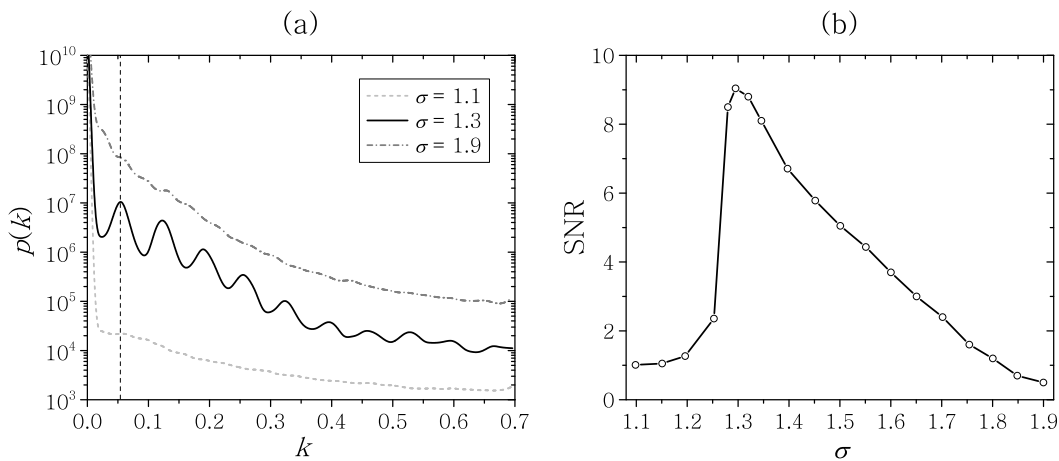}
\caption{Evidences for spatial coherence resonance in the studied HH medium. (a) The circular averages of structure functions presented in Fig.~\ref{fig:fig2}. The dashed vertical line at $k=k_{max}$ marks the spatial frequency of excitatory events that is resonantly enhanced by an intermediate level of additive Gaussian noise. (b) The signal-to-noise (SNR) ratio in dependence on $\sigma$ (the curve is just a guide to the eye).}
\label{fig:fig3}
\end{figure*}

Next, we exploit the circular symmetry of $P(k_x,k_y)$ as proposed by Carrillo et al. \cite{Car_2004} to obtain an estimate for the signal-to-noise ratio (SNR) of the spatial dynamics of excitatory events for different $\sigma$. In particular, we calculate the circular average of the structure function according to the equation
\begin{equation}
p(k)=\int_{\Omega_k}P(\vec{k})d\Omega_k,
\label{eq:eq5}
\end{equation}
where $\vec{k}=(k_x,k_y)$, and $\Omega_k$ is a circular shell of radius $k=|\vec{k}| $. Figure~\ref{fig:fig3}(a) shows $p(k)$ for the three different $\sigma$ corresponding to the values used already in Figs.~\ref{fig:fig1} and \ref{fig:fig2}. The presented results reveal that there indeed exists a particular spatial frequency, marked with the vertical dashed line at $k=k_{max}$, that is resonantly enhanced for some intermediate level of additive Gaussian noise. Note that subsequent local maxima are just higher harmonics of $k_{max}$. To quantify the ability of each $\sigma$ to extract the characteristic spatial frequency of the medium more precisely, we calculate the SNR as the peak height at $k_{max}$ normalized with the background fluctuations in the system; namely
$p(k_{max})/\tilde{p}$, where $\tilde{p}=[p(k_{max}-\Delta k_a)+p(k_{max}+\Delta k_b)]/2$ is an approximation for the level of background fluctuations in the system, whereby $\Delta k_a$ and
$\Delta k_b$ mark the estimated width of the peak around $k_{max}$ at the optimal $\sigma$. This is the spatial counterpart of the measure frequently used for quantifying constructive effects of noise in the temporal domain of dynamical systems, \cite{Gam_1998} whereas a similar measure for quantifying effects of noise on the spatial dynamics of spatially extended systems was also used in. \cite{Car_2004} Figure~\ref{fig:fig3}(b) shows how the SNR varies with $\sigma$. It is evident that there exists an optimal level of additive Gaussian noise for which the peak of the circularly averaged structure function is best resolved, thus clearly indicating the existence of spatial coherence resonance in the diffusively coupled HH medium.

We argue that the existence of spatial coherence resonance in the studied HH medium, and with it related remarkable order of the spatial dynamics of noise-induced excitatory events observed by intermediate $\sigma$, must be primarily attributed to the characteristic noise-robust excursion time inherent to all excitable systems. \cite{Pik_1997} The noise-robust excursion time, together with the rate of diffusive spread that is proportional to $\sqrt{D}$, introduces an eigenfrequency of spatial waves that can be resonantly enhanced by an appropriate intensity of noise, as recently shown analytically for a simple toy model of excitable dynamics. \cite{Mm_2006} Our results thus extend the scope of spatial coherence resonance by showing that real-life motivated more complex models of excitable dynamics, such as the HH mathematical model, \cite{Hod_1952} fully conform to findings obtained on simpler models, and that indeed this phenomenon may have important implications in several biological systems.

\section{Decoherence of spatial dynamics by small-world interactions}
\begin{figure*}
\centering
\includegraphics[width=14cm]{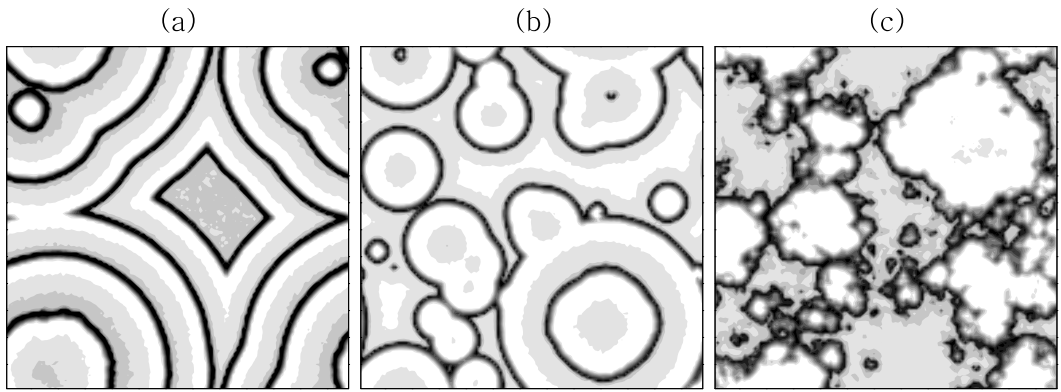}
\caption{Characteristic snapshots of the spatial profile of $V_{i,j}$ obtained at the optimal value of $\sigma=1.3$ for (a) $q=0.0001$, (b) $q=0.0008$, and (c) $q=0.002$.}
\label{fig:fig4}
\end{figure*}

\begin{figure*}
\centering
\includegraphics[width=14cm]{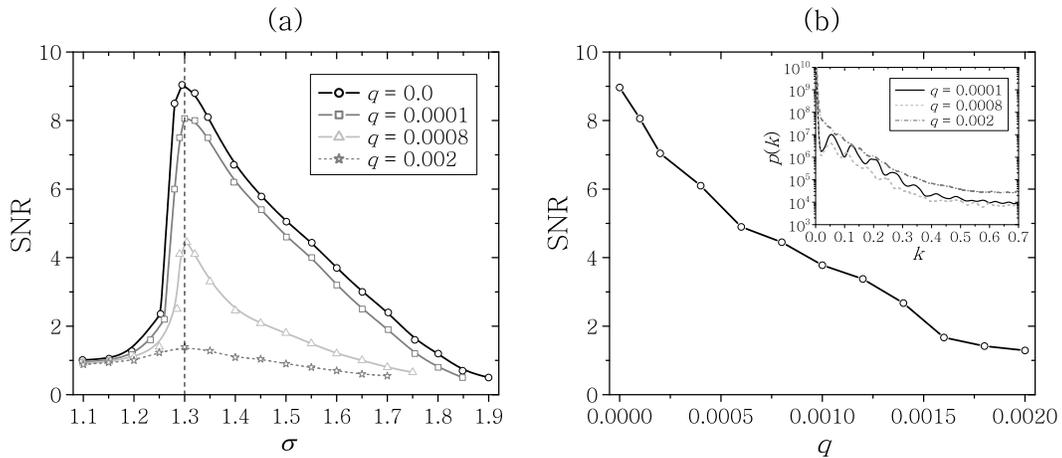}
\caption{Evidences for decoherence of the spatial dynamics in the studied HH medium with small-world topology. (a) The signal-to-noise ratio (SNR) in dependence on $\sigma$ for different $q$. (b) SNR in dependence on $q$ by the optimal $\sigma$ denoted by a dashed vertical line in panel (a). Inset features the circular average of the structure function $p(k)$ for the optimal $\sigma$ and three different values of $q$, corresponding to the three snapshots presented in Fig.~\ref{fig:fig4}. The values depicted in the main panels resulted from averaging the results over 30 different realizations of the interaction network by each $q$, whereas curves are just guides to the eye.}
\label{fig:fig5}
\end{figure*}

Next, it is of interest to investigate what are the impacts of small-world topology on the above-described phenomenon of spatial coherence resonance in the diffusively coupled HH medium. For this purpose we set $q>0$ and examine three characteristic snapshots of $V_{i,j}$ obtained by $\sigma=1.3$, which according to the results presented in Section III is the optimal noise intensity warranting superbly ordered spatial dynamics of excitatory events. Noteworthy, since we consider regular small-world networks warranting an equal number of inputs ($z=4$) to each coupled neuron, the optimal $\sigma$ for coherent pattern formation (if at all possible) does not depend on $q$, as will be confirmed also in Fig.~\ref{fig:fig5}(a) below. The snapshots presented in Fig.~\ref{fig:fig4} clearly evidence that increasing values of $q$ progressively hinder coherent pattern formation. In particular, while for $q=0.0001$ [Fig.~\ref{fig:fig4}(a)] the spatial periodicity of excitatory waves is still eligible, the same is much less true for $q=0.0008$ [Fig.~\ref{fig:fig4}(b)], and is certainly completely false for $q=0.002$ [Fig.~\ref{fig:fig4}(c)]. Thus, if as little as $0.08$ \% of nearest-neighbor links are rewired the pattern formation is impaired, while by $\approx 0.2$ \% noisy perturbations are completely unable to induce coherent spatial dynamics of excitatory events. Indeed, the snapshot presented in Fig.~\ref{fig:fig4}(c) lacks any ordered structure, although the same intensity of noise applied to a diffusively coupled HH medium evokes superbly ordered circular waves [see Fig.~\ref{fig:fig1}(b)] evidencing fascinating examples of spatial pattern formation out of noise.

To give a quantitative analysis of the observed phenomenon outlined in Fig.~\ref{fig:fig4}, we calculate the circularly average structure function $p(k)$ of the spatial dynamics in dependence on $\sigma$ and $q$. The inset in Fig.~\ref{fig:fig5}(b) shows three $p(k)$ obtained by the same parameter values yielding snapshots of spatial profiles shown in Fig.~\ref{fig:fig4}. In accordance with the interpretation of results presented in Section III, it can be inferred that increasing values of $q$ indeed impair the ability of noise to evoke ordered patterns with a predominant spatial frequency, because the peak height of $p(k)$ at $k=k_{max}$ becomes smaller in comparison to the level of background noise as the fraction of rewired links increases. While by $q=0.0001$ and $q=0.0008$ the predominant spatial periodicity is still present, it vanishes completely by $q=0.002$ as the outlay of $p(k)$ becomes essentially flat. These observations can be captured succinctly by calculating the SNR in dependence on $\sigma$ by different $q$. The results presented in Fig.~\ref{fig:fig5}(a) evidence clearly that the phenomenon of spatial coherence resonance fades continuously as $q$ increases, disappearing almost completely by $q=0.002$. As already noted, however, the optimal $\sigma$ for coherent pattern formation (as much as allowed by the topology of the underlying interaction network) remains virtually identical by all $q$, equaling $\sigma=1.3$, as denoted by the dashed vertical line in Fig.~\ref{fig:fig5}(a). We exploit this fact in Fig.~\ref{fig:fig5}(b), where the SNR is plotted in dependence on $q$ for the optimal $\sigma$. Clearly, the SNR decreases continuously as $q$ increases, thus evidencing the decoherence of noise-induced spatial dynamics due to the introduction of small-world topology in the studied HH spatially extended system.

While the existence of spatial coherence resonance in the diffusively coupled HH medium was attributed to the characteristic noise-robust excursion time of the local excitable dynamics, \cite{Pik_1997} which together with the rate of diffusive spreading proportional to $\sqrt{D}$ introduces an eigenfrequency, or characteristic spatial scale, of waves that can be resonantly enhanced by an appropriate intensity of noise, \cite{Mm_2006} the decoherence of the spatial dynamics due to the introduction of small-world topology is simply a consequence of the disruption of this inherent spatial scale. \cite{Perc_2005} In particular, while the excursion time of individual space units remains unaltered by the introduction of long-range couplings, the spread rate of excitations is indirectly very much affected by increasing values of $q$. Namely, the introduction of shortcut links decreases the typical path length between two arbitrary sites in comparison to a diffusively coupled network. Thus, while the spread rate of excitations is still proportional to $\sqrt{D}$, the typical path length between two arbitrary grid units decreases dramatically, which in turn has the same effect as if $D$ would increase. Therefore, in a small-world network, a locally induced excitation can instantly reach much more distant sites than in case of strict diffusive coupling, in turn facilitating noise-induced synchronization of distinct network units, \cite{Gao_2001} or assuring fast response abilities of the system. \cite{Lago_2000} Importantly, however, the typical path length between two arbitrary sites decreases only on average, meaning that there does not exist an exact path length defining the distance between all possible pairs of sites in a small-world network. Thus, a given local excitation can, during the excursion time, propagate to the most distant site or just to its nearest neighbor, whereby clearly the well-defined inherent spatial scale existing for the diffusive coupling is lacking, and this leads to spatial decoherence of noise-induced spatial patterns already for very small values of $q$, as emphasized throughout this Section.

\section{Summary}
We have shown that temporally and spatially uncorrelated additive Gaussian noise is able to extract a characteristic spatial frequency of an excitable HH medium in a resonant manner if the latter is coupled diffusively. In particular, it then exists an optimal level of additive noise for which the spatial dynamics of the system is highly coherent and superbly ordered. Thereby, no additional deterministic inputs were introduced to the system and all units were initiated from steady-state excitable conditions. Thus, the results presented in Section III offer convincing evidences for the existence of spatial coherence resonance in the diffusively coupled HH medium. On the other hand, when the interactions amongst coupled neurons are governed by a small-world network, the ability of noise to induce ordered spatial dynamics vanishes quickly as $q$ increases. Our results suggest that as little as $0.2$ \% of rewired nearest-neighbor links suffice to completely hinder coherent pattern formation out of noise within the currently employed setup. Noteworthy, we have performed additional simulations using the FitzHugh-Nagumo system \cite{Fh1961} and the Goldbeter model for calcium oscillations \cite{Gold1990} as governing units of the medium, whereby the former excitable dynamics is governed by a steady node while the latter is characterized by a stable focus situated prior to a Hopf bifurcation. Irrespective of these particularities, we were able to observe coherent patterns emerging out of noise provided the units were diffusively coupled. Conversely, if small-world connectivity was introduced the ordered waves vanished quickly already by $q \ll 1$, albeit stable foci could sustain some order in the spatial dynamics up to $0.5$ \% of rewired links due to larger re-settlement times associated with their excitable dynamics. Larger re-settlement times following an excitation yield waves with larger wavelengths, and hence then the probability of shortcuts to link two quiescent units increases, thus leaving the spatial dynamics temporarily unaffected. Aside from these rather minute differences, however, we conclude that the present findings appear to be robust with respect to variations in models governing the local excitable dynamics of diffusive and small-world networks.

Since the present results rely on the use of the HH model for the description of the local dynamics of the excitable medium, the present study might also have some biological implications. Specifically for neural systems, it has been argued that excitable media guarantee robust signal propagation through the tissue in a substantially noisy environment. \cite{Izy_2000} It would thus be fascinating to elucidate if spatial coherence resonance in the neural system can be confirmed also experimentally. The above theoretical results indicate that such findings might indeed be attainable at least if the units are diffusively coupled, and that it thus seems reasonable to pursue this problem in the future. Moreover, our results obtained when using small-world networks may help by revealing structural functionality of complex topologies within the neural apparatus, although additional studies regarding the concept of function-follow-form \cite{Towle_1999, Hilgetag_2000, Segev_2003, Volman_2005} in the presence of noise and other uncertainties are necessary to clarify the importance of different structural physiological properties of such networks.

\begin{acknowledgments}
X. Sun and Q. Lu are grateful for the support of the National Science Foundation of China (Grants 10432010 and 10702023). M. Perc acknowledges support from the Slovenian Research Agency (Grant Z1-9629).
\end{acknowledgments}

\end{document}